\documentclass[conference]{IEEEtran}
\IEEEoverridecommandlockouts
\usepackage{cite}
\usepackage{amsmath,amssymb,amsfonts}
\usepackage{algorithmic}
\usepackage{graphicx}
\usepackage{textcomp}
\usepackage{xcolor}
\usepackage{acronym}
\usepackage{siunitx}
\usepackage{soul}
\acrodef{sps}[SPS]{Shipboard Power System}
\acrodef{ess}[ESS]{Energy Storage System}
\acrodef{pcs}[PCS]{Power Conversion System}
\acrodef{mmc}[MMC]{Modular Multilevel Converter}
\acrodef{sl}[SL]{Special Load}
\acrodef{afe}[AFE]{Active Front End}
\acrodef{bdc}[BDC]{Bidirectional Buck-Boost Converter}
\acrodef{esr}[ESR]{Equivalent Series Resistance}
\acrodef{esl}[ESL]{Equivalent Series Inductance}

\def\BibTeX{{\rm B\kern-.05em{\sc i\kern-.025em b}\kern-.08em
    T\kern-.1667em\lower.7ex\hbox{E}\kern-.125emX}}
\begin{document}

\title{Short-Circuit Current Calculation in DC Shipboard Microgrid Using Operational Admittances

}

\author{\IEEEauthorblockN{Pietro Cepollini}
\IEEEauthorblockA{\textit{University of Genova - DITEN} \\
pietro.cepollini@edu.unige.it}
\and
\IEEEauthorblockN{Fabio D'Agostino}
\IEEEauthorblockA{\textit{University of Genova - DITEN} \\
fabio.dagostino@unige.it}
\and
\IEEEauthorblockN{Federico Silvestro}
\IEEEauthorblockA{\textit{University of Genova - DITEN} \\
federico.silvestro@unige.it}
\and
\IEEEauthorblockN{Antonio Chiarelli}
\IEEEauthorblockA{\textit{Fincantieri S.p.A.} \\
antonio.chiarelli@fincantieri.it}
}

\maketitle

\begin{abstract}

The increasing adoption of DC shipboard power systems is driven by the need to integrate energy storage systems seamlessly onboard. In this context, accurate calculation of short-circuit current transients is critical for sizing fast-acting protection devices.   
This paper proposes a methodology to calculate short-circuit current transients based on operational admittance modelling of network elements and power converters. The superposition principle is then applied to calculate current contributions and voltage time profiles.
The methodology is validated with a detailed EMT model of a notional LVDC shipboard microgrid section consisting of an active front end, a bidirectional buck-boost converter and two special loads. Results demonstrate that the methodology accurately tracks nodal voltages and the fault current contributions dynamics up to the prospective fault current peak, thus providing a computationally efficient tool for fault analysis and protection coordination in LVDC microgrids.
\end{abstract}

\begin{IEEEkeywords}
DC Microgrids, Shipboard Power Systems, Short-circuit calculation, Operational Admittance.
\end{IEEEkeywords}

\section{Introduction}

DC \acp{sps} have become a trend in recent years due to the need to easily integrate \acp{ess}.

Although some DC \acp{sps} integrate \acp{ess} directly connected to the \ac{sps} \cite{SiemensFloating}, the majority of the solutions in industry \cite{ABBControlledc}, recommended practices \cite{IEEE451}, and literature \cite{DCRev} employ \acp{pcs} to integrate the \acp{ess}. Such architecture decouples \ac{sps} and \ac{ess} voltage levels, allows \ac{ess} current control and limitation and permits the integration of hierarchical control of the resource.

In this context, the prospective short-circuit evaluation is a critical topic.
This current is dominated by DC-link capacitors discharge from \acp{pcs} capacitive filters and additional capacitor banks integrated to reduce the magnitude of voltage oscillations induced by \acp{sl} \cite{PPLCapDC}.
Additionally, the peak magnitude and its transient evolution are dictated by the network passive parameters, as the cable and busbar impedances \cite{Sivori}.
Moreover, LVDC \acp{sps} typically operate as ungrounded (IT) systems, where the fault current resulting from a single pole-to-ground event is inherently limited and mainly causes a pole-voltage displacement, requiring insulation monitoring devices \cite{DoerryGrounding2025}. Consequently, pole-to-pole short-circuit transients represent the most severe fault condition, effectively dictating the requirements for the design and sizing of protective devices \cite{FootPrintePOLEPOLE}.

DC short-circuit current exhibits two main phases: a rapid capacitive discharge peak and a slower contribution from the active sources. Accurately evaluating the peak both in magnitude and time is critical to determine electro-dynamical stresses, properly size fast-acting protection devices as solid state circuit breakers~\cite{ABB} and ensure their selective operation to prevent fault propagation~\cite{Fabrizio}.

Standard IEC 61660-1 provides a methodology to compute short-circuit currents in the presence of DC auxiliary systems in power plants and substations via standard approximation functions.
However, this methodology does not include modern \acp{pcs} as \acp{afe} or \acp{bdc}.

To address these limitations, authors in \cite{MeshDCFaults} expand the standard by integrating AC-DC connection interfaces and distributed energy resources, such as photovoltaic systems, into a generalized matrix formulation. Similarly, the work in \cite{SimplifiedShortCircuit} extends the standard by proposing a superposition method of time-approximated functions for meshed grids.
Alternatively, analytical time-domain approaches based on linearized state-space formulations have been recently proposed to evaluate the short-circuit current dynamics and assess solid state circuit breakers protection selectivity in LVDC \acp{sps}~\cite{Fabrizio}.

In contrast to these time-domain or approximated approaches, Laplace-domain and operational methodologies have emerged in the HVDC sector employing operational impedance to accurately calculate the fault transient.
Authors in \cite{PoleToPoleSC} estimate the short-circuit current dynamic after pole-to-pole fault via an RLC operational impedance approximation of \acp{mmc}. This approach enables the efficient computation of the short-circuit current dynamics up to the prospective peak.
Authors in \cite{MethodSCHVDCgrids} extend this concept by integrating traveling-wave transmission line models and \acp{mmc} control systems via the bus impedance transfer matrix.
However, these approaches are specifically tailored to HVDC transmission systems and \acp{mmc}. Their applicability to LVDC \acp{sps} remains unexplored, particularly considering the different converter configurations and shorter cable lengths that characterize shipboard grids.

To bridge this gap, the main contribution of this paper is a methodology to calculate pole-to-pole short-circuit current transients based on the operational admittance modelling of network elements and power converters. By applying the superposition principle, individual current contributions and nodal voltage time profiles are derived accurately up to the prospective fault current peak. 

The methodology is validated through a detailed EMT model of a notional LVDC shipboard microgrid section consisting of an active front end, a bidirectional converter, and specific loads. Results demonstrate the effectiveness to track nodal voltages and the fault current contributions dynamics up to the prospective fault current peak, thus providing a computationally efficient tool for fault analysis and protection coordination in LVDC microgrids.

\section{Operational Admittance Network Modelling}
The network includes the set of nodes $\mathcal{N}$, representing the main buses. 
Following the nodal injection convention (\textit{i.e.} positive values indicate injected currents), the dynamical relation between the nodal voltage vector $\mathbf{v}_\mathcal{N}(s)$ and the nodal current injection vector $\mathbf{i}_\mathcal{N}(s)$ is described by:
\begin{equation}
\mathbf{i}_\mathcal{N}(s)=\mathbf{Y}_\mathcal{N}(s)\mathbf{v}_\mathcal{N}(s),
\end{equation}
where $\mathbf{Y}_\mathcal{N}(s)$ denotes the operational nodal admittance matrix.

The interconnecting lines $\mathcal{L}$ are characterized by their operational series admittance, collected in the line operational internal admittance $\mathbf{y}_{\mathcal{L}}(s)$, while the line currents are grouped into $\mathbf {i}_{\mathcal {L}}(s)$.
Moreover, the topology of the grid is described by the network incidence matrix $\mathbf {A}_{(\mathcal {L},\mathcal {N})}$.

Therefore, the nodal admittance matrix can be derived as follows:
\begin{equation}
\label{eq:nodalAdmittance}
\mathbf {Y}_{\mathcal {N}}(s) = \mathbf {A}_{(\mathcal {L},\mathcal {N})}^{T}\mathbf {Y}_{\mathcal {L}}(s)\mathbf {A}_{(\mathcal {L},\mathcal {N})} + \mathbf {Y}_{\mathcal {T}}(s),
\end{equation}
where $\mathbf{Y}_{\mathcal{T}}(s) = \operatorname{diag}\!\big(\mathbf{y}_{\mathcal{T}}(s)\big)$ is the nodal shunt admittance operational matrix and $\mathbf{Y}_{\mathcal{L}}(s)=\operatorname{diag}\!\big(\mathbf{y}_{\mathcal{L}}(s)\big)$ is the line admittance operational matrix.

\subsection{Nodal elements}
\label{NodalElements}
The construction of $\mathbf{Y}_{\mathcal{T}}(s)$ depends on the characterization of the nodal loads and sources.

Specifically, the set of passive shunt elements $\mathcal{P}$ is described by the admittance vector $\mathbf{y}_{\mathcal{P}}(s)$, which aggregates all passive nodal shunt admittances.

The set of nodal generations $\mathcal{G}$ comprises the active sources, characterized by an internal impedance, and a pre-fault open-circuit voltage.
To facilitate their integration into the subsequent short-circuit calculations, they are represented by their Norton equivalent circuits. Accordingly, these sources are modeled using an internal admittance vector $\mathbf{y}_{\mathcal{G}}(s)$ and an equivalent current injection vector $\mathbf{i}_{\mathcal{G}}^{\mathrm{pre}}$.

Thus, the nodal shunt admittance matrix $\mathbf{Y}_{\mathcal{T}}(s)$ can be derived by summing the passive and active contributions:
\begin{equation}
\mathbf{Y}_{\mathcal{T}}(s) = \mathbf{Y}_{\mathcal{P}}(s) + \mathbf{Y}_{\mathcal{G}}(s),
\end{equation}
where $\mathbf{Y}_{\mathcal{P}}(s) = \operatorname{diag}\!\big(\mathbf{y}_{\mathcal{P}}(s)\big)$ is the  passive shunt operational admittance matrix and $\mathbf{Y}_{\mathcal{G}}(s) = \operatorname{diag}\!\big(\mathbf{y}_{\mathcal{G}}(s)\big)$ is the source internal operational admittance matrix.

Finally, both the set of constant current loads $\mathcal{D}$ and the constant current sources within the set $\mathcal{J}$ are modeled as ideal independent current sources. While they do not affect the nodal admittance matrix, their current injection values are collected into the demand vector $\mathbf{i}_{\mathcal{D}}^{\mathrm{pre}}$ and the source current vector $\mathbf{i}_{\mathcal{J}}^{\mathrm{pre}}$, respectively.

It is important to note that once the pre-fault nodal voltages $\mathbf{v}_\mathcal{N}^{\mathrm{pre}}$ and line currents $\mathbf{i}_\mathcal{L}^{\mathrm{pre}}$ are calculated (\textit{e.g.}, through a power flow calculation or an EMT simulation), the converter control strategies become secondary for the short-circuit transient analysis. 
Therefore, $\mathbf{i}_{\mathcal{G}}^{\mathrm{pre}}$, $\mathbf{i}_{\mathcal{J}}^{\mathrm{pre}}$ and  $\mathbf{i}_{\mathcal{D}}^{\mathrm{pre}}$ do not need to be specified, as $\mathbf{v}_\mathcal{N}^{\mathrm{pre}}$ and $\mathbf{i}_\mathcal{L}^{\mathrm{pre}}$ are the only pre-fault quantities entering the formulation.

\subsection{Short-circuit calculation methodology}
Given that the short-circuit operational admittance network modelling is linear, superposition applies. The post-fault state is expressed by combining the steady-state pre-fault conditions and the variations introduced by the fault:

\begin{equation}
\begin{cases}
\mathbf{v}_\mathcal{N}^{\mathrm{post}}(t)=\mathbf{v}_\mathcal{N}^{\mathrm{pre}}+\Delta\mathbf{v}_\mathcal{N}(t)\\
\mathbf{i}_\mathcal{N}^{\mathrm{post}}(t)=\mathbf{i}_\mathcal{N}^{\mathrm{pre}}+\Delta\mathbf{i}_\mathcal{N}(t)
\end{cases},
\end{equation}
where $\mathbf{v}_\mathcal{N}^{\mathrm{pre}}$ and $\mathbf{i}_\mathcal{N}^{\mathrm{pre}}$ are the pre-fault nodal voltages and current injections, $\Delta\mathbf{v}_\mathcal{N}(t)$ and $\Delta\mathbf{i}_\mathcal{N}(t)$ are the transient variations induced by the fault, while $\mathbf{v}_\mathcal{N}^{\mathrm{post}}(t)$ and $\mathbf{i}_\mathcal{N}^{\mathrm{post}}(t)$ represent the resulting post-fault nodal voltage and current injection vectors.

The previous equations are expressed in the Laplace-domain to implement the previously derived operational matrices:
\begin{equation}
\label{eq:superpositionLaplace}
\begin{cases}
\mathbf{v}_\mathcal{N}^{\mathrm{post}}(s)=\frac{\mathbf{v}_\mathcal{N}^{\mathrm{pre}}}{s}+\Delta\mathbf{v}_\mathcal{N}(s)\\
\mathbf{i}_\mathcal{N}^{\mathrm{post}}(s)=\frac{\mathbf{i}_\mathcal{N}^{\mathrm{pre}}}{s}+\Delta\mathbf{i}_\mathcal{N}(s)
\end{cases}.
\end{equation}

To evaluate the nodal voltage transient, the operational nodal impedance matrix $\mathbf{Z}_\mathcal{N}(s) = \mathbf{Y}^{-1}_\mathcal{N}(s)$ is introduced to link voltage and current variations:
\begin{equation}
\Delta\mathbf{v}_\mathcal{N}(s) = \mathbf{Z}_\mathcal{N}(s) \Delta\mathbf{i}_\mathcal{N}(s).
\end{equation}

Considering a single fault at the $k$-th node with fault impedance $Z_f(s)$, the nodal current variation vector reduces to:
\begin{equation}
  \Delta\mathbf{i}_\mathcal{N}(s) =-i_f(s)\mathbf{e}_k =[0, \dots, -i_f(s), \dots, 0]^{T},
\end{equation}
where $i_f(s)$ represents the fault current and $\mathbf{e}_k$ is the $k$-th standard basis vector, \textit{i.e.} an $\mathcal{N}$ dimension zero column vector with the k-th element equal to one.
This leads to:
\begin{equation}
\label{eq:VoltageVariationAfterFault}
\Delta\mathbf{v}_\mathcal{N}(s) = -\mathbf{Z}_\mathcal{N}(s) \mathbf{e}_k i_f(s).
\end{equation}
Consequently, the post-fault voltage at the faulted point must satisfy $\mathbf{e}_k^T \mathbf{v}_\mathcal{N}^{\mathrm{post}}(s) = Z_f(s)i_f(s)$. Thus, combining \eqref{eq:superpositionLaplace} and \eqref{eq:VoltageVariationAfterFault}, the following relation can be derived:
\begin{equation}
Z_f(s)i_f(s) = \frac{\mathbf{e}_k^T \mathbf{v}_\mathcal{N}^{\mathrm{pre}}}{s} - \left( \mathbf{e}_k^T \mathbf{Z}_\mathcal{N}(s) \mathbf{e}_k \right) i_f(s).
\end{equation}
Therefore, the fault current can be derived as:
\begin{equation}
\label{eq:FaultCurrentMatrix}
i_f(s) = \frac{1}{s}\frac{\mathbf{e}_k^T \mathbf{v}_\mathcal{N}^{\mathrm{pre}}}{ \mathbf{e}_k^T \mathbf{Z}_\mathcal{N}(s) \mathbf{e}_k + Z_f(s)}.
\end{equation}
Thus, combining \eqref{eq:FaultCurrentMatrix}, \eqref{eq:VoltageVariationAfterFault} and \eqref{eq:superpositionLaplace} the post-fault voltages are derived as:
\begin{equation}
\label{eq:nodal_voltages}
\mathbf{v}_\mathcal{N}^{\mathrm{post}}(s)=\frac{1}{s}\left[\mathbf{v}_\mathcal{N}^{\mathrm{pre}}-\frac{\mathbf{Z}_\mathcal{N}(s)\mathbf{e}_k \left(\mathbf{e}_k^T\mathbf{v}_\mathcal{N}^{\mathrm{pre}}\right)}{\mathbf{e}_k^T\mathbf{Z}_\mathcal{N}(s)\mathbf{e}_k+Z_f(s)}\right].
\end{equation}
It is worth noting that the term $\mathbf{Z}_\mathcal{N}(s)\mathbf{e}_k = \mathbf{Y}_\mathcal{N}^{-1}(s)\mathbf{e}_k=\mathbf{z}_{col,k}(s)$ represents the $k$-th column of the nodal impedance matrix. This vector can be computed by solving the linear system $\mathbf{Y}_\mathcal{N}(s)\mathbf{z}_{col,k}(s) =\mathbf{e}_k$, avoiding a full matrix inversion.

Furthermore, the line currents are determined by expressing the current variations as a function of the voltage variations: 
\begin{equation}
    \Delta\mathbf{i}_\mathcal{L}(s)=\mathbf{Y}_{\mathcal {L}}(s)\mathbf {A}_{(\mathcal {L},\mathcal {N})}\Delta\mathbf{v}_\mathcal{N}(s).
    \end{equation}
Moreover, the post-fault line current vector can be expressed as:
\begin{equation}
\label{eq:line_currents}
\mathbf{i}_\mathcal{L}^{\mathrm{post}}(s)=\frac{1}{s}\left[\mathbf{i}_\mathcal{L}^{\mathrm{pre}}-\mathbf{Y}_{\mathcal {L}}(s)\mathbf {A}_{(\mathcal {L},\mathcal {N})}\frac{\mathbf{Z}_\mathcal{N}(s)\mathbf{e}_k \left(\mathbf{e}_k^T\mathbf{v}_\mathcal{N}^{\mathrm{pre}}\right)}{\mathbf{e}_k^T\mathbf{Z}_\mathcal{N}(s)\mathbf{e}_k+Z_f(s)}\right].
\end{equation}

The time domain post-fault values of $\mathbf{v}_\mathcal{N}^{\mathrm{post}}$ and  $\mathbf{i}_\mathcal{L}^{\mathrm{post}}$ are derived by recognizing that the $1/s$ term in \eqref{eq:nodal_voltages} and \eqref{eq:line_currents} corresponds to a step excitation. Consequently, the time response is computed using the \texttt{step} command in MATLAB applied to the resulting transfer functions.

\section{Study case}
\begin{figure}[htbp]
    \centering
\includegraphics[width=\columnwidth]{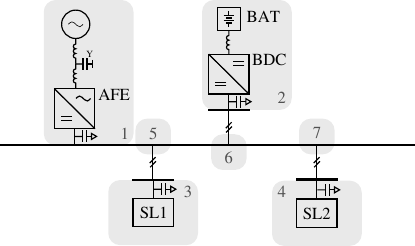}
    \caption{One line diagram of the selected case study. Triangle symbols denote the connection to the negative pole.}
    \label{fig:CaseStudy}
\end{figure}

To validate the proposed methodology, a case study based on the mixed MVAC-LVAC radial \ac{sps} topology from \cite{Architettura} is presented. The original architecture comprises two independent AC power plants and distribution networks. From this, two distinct DC subsections are derived to facilitate the integration of \acp{sl} and \acp{ess} at a nominal voltage of \SI{1000}{\volt}. This study focuses on a single DC subsection, as depicted in Figure~\ref{fig:CaseStudy}.

\subsection{System architecture and control}
The DC grid is interfaced with the AC distribution via a \SI{4}{\mega\watt} \ac{afe}, which employs an inner AC \emph{dq} current control loop and an outer DC voltage control loop. The \ac{ess} consists of an \SI{800}{\volt} nominal battery pack, connected to the DC bus through a \SI{2}{\mega\watt} \ac{bdc}. The \ac{bdc} relies on an inner battery current loop and an outer voltage loop. Both converters feature a single-stage non-isolated topology.

To ensure proportional power sharing, both active converters implement a droop control strategy. Additionally, both units feature an under-voltage protection mechanism that disables the switching devices if the bus voltage drops below \SI{500}{\volt}.
The system supplies two \acp{sl}, each rated at \SI{1}{\mega\watt}, which are assumed to operate at maximum power during the pre-fault steady state.

\subsection{Network parameters and physical layout}
Regarding the physical layout, the \ac{afe} is directly integrated into the main switchboard. 

In contrast, the \ac{bdc} and the \acp{sl} are connected via dedicated cables.
Each interconnecting line, as well as the switchboard busbar sections, is characterized by a lumped-parameter series RL equivalent circuit. 

To accurately capture the transient behavior during the short-circuit discharge, each capacitor bank is characterized by accounting for both its parasitic \ac{esr} and \ac{esl}. 
These parameters, along with the total capacitance, are derived by scaling commercial \SI{950}{\micro\farad} capacitor parameters \cite{KEMET_C44U_M}, while the \ac{esl} is attributed mainly due to stray effects and set to typical values \cite{DELLACASA}. 
Table~\ref{tab:system_parameters} summarizes the cable system, busbar and capacitor parameters.

\begin{table}[t]
    \caption{System Parameters: DC-Link Capacitors and Cables}
    \label{tab:system_parameters}
    \centering
    \footnotesize
    \renewcommand{\arraystretch}{1.4}
    \setlength{\tabcolsep}{4pt} 
    \begin{tabular}{l c c c | l c c}
        \hline
        \textbf{Comp.} & \textbf{C} [mF] & \textbf{ESR} [\si{\micro\ohm}] & \textbf{ESL} [nH] & \textbf{Line} & \textbf{R} [m\si{\ohm}] & \textbf{L} [\si{\micro H}] \\
        \hline
        AFE  & 39.9 & 27.38 & 30 & Busbar & 0.015 & 0.45 \\
        BDC  & 16.1 & 67.65 & 30 & BDC    & 1.16  & 15.6 \\
        SL1 & 32.3 & 33.82 & 30 & SL1   & 5.40  & 46.1 \\
        SL2 & 32.3 & 33.82 & 30 & SL2   & 2.70  & 23.1 \\
        \hline
    \end{tabular}
\end{table}

\subsection{Component Operational Admittance Characterization}
\begin{figure}[htbp]
    \centering
\includegraphics[width=\columnwidth]{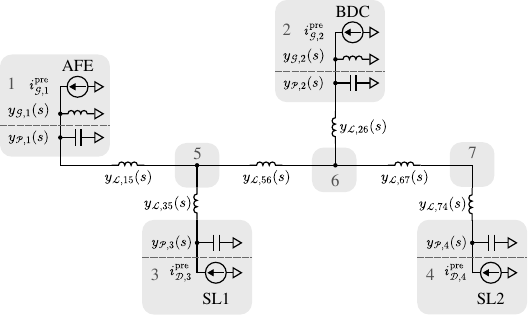}
    \caption{Operational admittance-based short-circuit equivalent circuit.}
    \label{fig:CaseStudyAdmi}
\end{figure}

To evaluate the system behavior during the fault transient, the equivalent short-circuit model is thus derived as shown in Fig.~\ref{fig:CaseStudyAdmi}.

The \ac{afe} and \ac{bdc} are modeled as operational Norton equivalents, with $\mathbf{i}_{\mathcal{G}}$ and $\mathbf{y}_{\mathcal{G}}(s)$.
During the first instant of the fault transient, the n-th converter short-circuit is limited by its choke filter inductance $L_{f,n}$. Therefore the converter internal admittance is derived as:
\begin{equation}
\label{eq:choke}
 y_{\mathcal{G},n}(s)= \frac{1}{sL_{f,n}}.
\end{equation}

This modelling choice implies that, immediately after the fault, the converter current rises with a derivative determined by the difference between the pre-fault network and the post-fault nodal voltage divided by the filter inductance. 

In reality, the current rise is determined by the difference between the source and the post-fault voltage divided by the filter inductance. Nevertheless, this approach proves to be conservative since the pre-fault network voltage, due to the boost operation, is always higher than the source voltage.
Line admittances are expressed as: 
\begin{equation}
    y_{\mathcal{L}}(s) = \frac{1}{R_{c} + s L_{c}},
\end{equation}
where $R_{c}$ and $L_{c}$ represent the equivalent line resistance and inductance.
Moreover, \acp{sl} are modelled as constant current loads $\mathbf{i}_{\mathcal{D}}$.

Capacitors are characterized as passive shunt operation admittances $\mathbf{y}_{\mathcal{P}}(s)$:
\begin{equation}
y_{\mathcal{P}}(s) = \frac{1}{\mathrm{ESR} + s \mathrm{ESL} + \frac{1}{sC}},
\end{equation}
which can be reformulated as:
\begin{equation}
y_{\mathcal{P}}(s) = sC\frac{\omega_{n}^2}{s^2 + 2\zeta\omega_{n} s + \omega_{n}^2},
\end{equation}
where the capacitor undamped natural frequency is:
\begin{equation}
\omega_{n} = \frac{1}{\sqrt{\mathrm{ESL} \, C}},
\end{equation}
and the capacitor damping ratio is:
\begin{equation}
\zeta = \frac{\mathrm{ESR}}{2}\sqrt{\frac{C}{\mathrm{ESL}}}.
\end{equation}

\subsection{Methodology Validation}
The results of the proposed methodology are compared to a full EMT model developed in MATLAB/Simulink using the Simscape Specialized Power Systems blockset. Both \ac{afe} and \ac{bdc} are implemented as their ideal switching model. Both converters operate at a switching frequency equal to \SI{5}{\kilo\hertz}. 
The employed digital control system is synchronized with the carrier signal, resulting in a single-sampling controller for the \ac{bdc} and a double-sampling controller for the \ac{afe}. 

\textcolor{black}{The ideal switching converter models are characterized by instantaneous commutation events which, interacting with the capacitor ESL and the converter snubber circuits, introduce high-frequency voltage spikes in the EMT results. Their amplitude and frequency content are not accurate, as their reproduction would require device-level semiconductor models, and they have a negligible influence on the fault transient under investigation.}

Three fault locations along the switchboard are considered, at nodes 5, 6 and 7. In all cases the fault impedance $Z_f(s)$ is assumed to be purely resistive and equal to \SI{1}{\milli\ohm}. For each location, the total short-circuit current, the current contributions of \ac{afe}, \ac{bdc} and \acp{sl}, and the respective nodal voltages are reported.

\begin{figure}[!t]
    \centering    \includegraphics[width=\columnwidth]{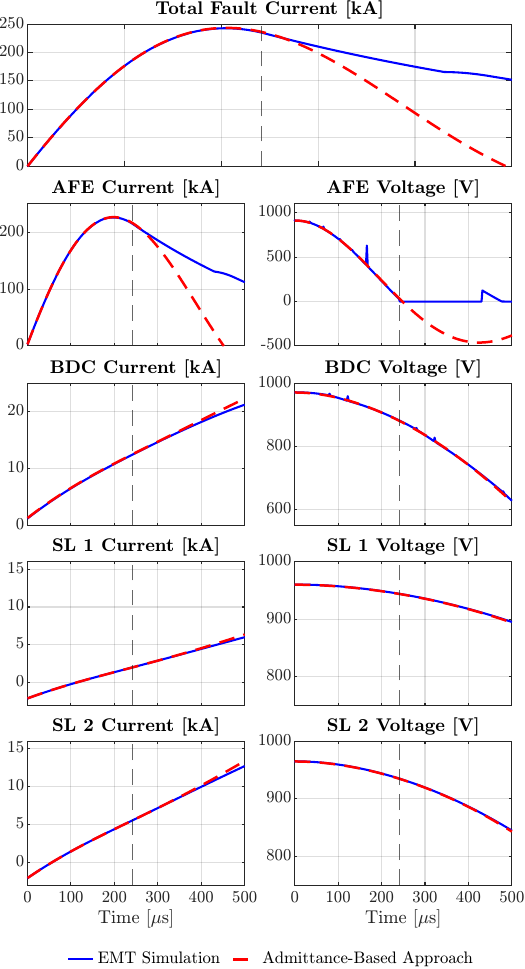}
    \caption{{Sources and loads short-circuit branch currents and nodal voltages after fault at node 5, $Z_f=$\SI{1}{\milli\ohm}. The dashed vertical line indicates the moment when the \ac{afe} diodes start freewheeling.}}
    \label{fig:SC5}
\end{figure}

Figure~\ref{fig:SC5} presents the system response to a fault at node 5, where the prospective peak reaches \SI{242.7}{\kilo\ampere} at \SI{204}{\micro\second}. The \ac{afe} contributes \SI{225.7}{\kilo\ampere} to the peak current, accounting for \SI{93}{\percent} of the total.

\begin{figure}[!t]
    \centering    \includegraphics[width=\columnwidth]{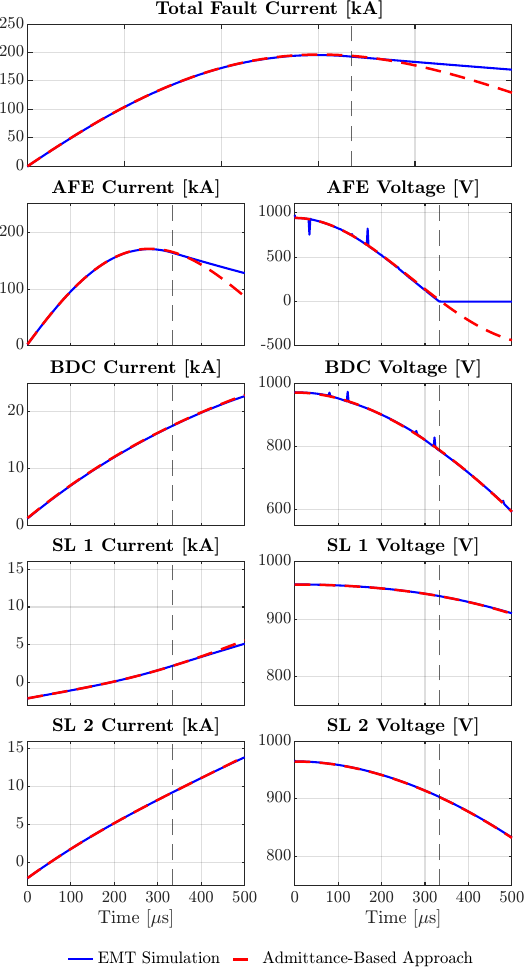}
    \caption{Sources and loads short-circuit branch currents and nodal voltages after fault at node 6, $Z_f=$\SI{1}{\milli\ohm}. The dashed vertical line indicates the moment when the \ac{afe} diodes start freewheeling.}
    \label{fig:Shortcircuitcurrent}
\end{figure}

For the fault at node 6, shown in Figure~\ref{fig:Shortcircuitcurrent}, the peak decreases to \SI{195.5}{\kilo\ampere} and is reached at \SI{298}{\micro\second}, with the \ac{afe} contribution dropping to \SI{86.7}{\percent} (\SI{169.5}{\kilo\ampere}).

\begin{figure}[!t]
    \centering
    \includegraphics[width=\columnwidth]{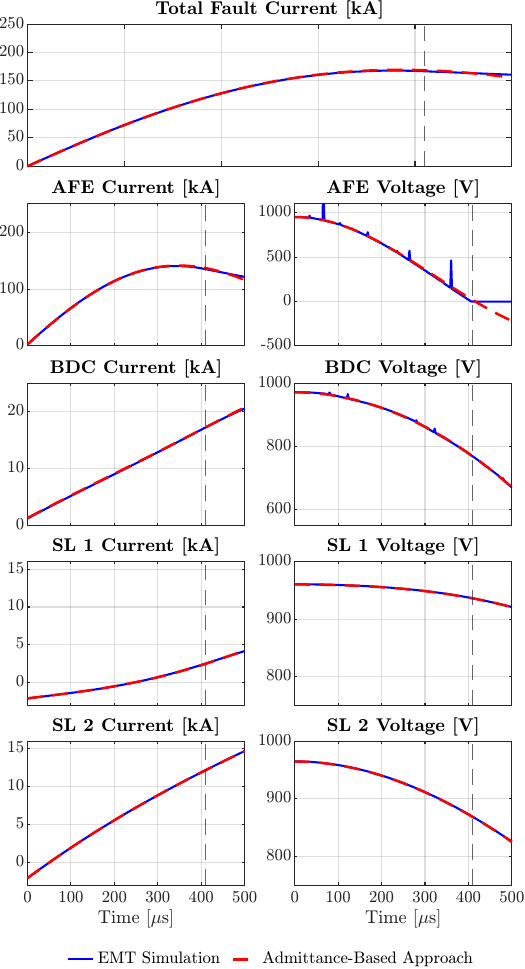}
    \caption{{Sources and loads short-circuit branch currents and nodal voltages after fault at node 7, $Z_f=$\SI{1}{\milli\ohm}. The dashed vertical line indicates the moment when the \ac{afe} diodes start freewheeling.}}
    \label{fig:SC7}
\end{figure}

Finally, Figure~\ref{fig:SC7} reports the fault at node 7, where the peak further decreases to \SI{168.3}{\kilo\ampere} at \SI{378}{\micro\second}. The \ac{afe} provides \SI{139.1}{\kilo\ampere}, equal to \SI{82.7}{\percent} of the total current.

The results show that the prospective peak decreases and occurs later as the fault location moves away from the \ac{afe}, whose \mbox{DC-link} capacitor is the largest of the system and, being connected directly to the switchboard with no additional current rise limiting elements, provides the dominant contribution in all the considered cases.

In all three cases, at \SI{242}{\micro\second}, \SI{334}{\micro\second} and \SI{410}{\micro\second} respectively, the \ac{afe} voltage in the EMT model drops to zero and does not reverse, as the internal diodes provide a local freewheeling path for the current which may damage the converter. Conversely, the linear model, being underdamped, predicts the voltage reversal. Beyond this point, the proposed methodology loses accuracy, as this nonlinear effect is not accounted for. Nevertheless, up to this point, the methodology effectively tracks both the individual resources current contributions and the nodal voltages.
\section{Conclusion}
This paper presented a short-circuit-current calculation methodology for DC shipboard microgrids. Operational admittance formulation is employed to derive both line current and nodal voltage transients after the fault.
The methodology is validated with an EMT switching model of a notional shipboard LVDC section demonstrating the coherence of the methodology results up to the prospective current peak.
Future research will focus on the inclusion of cable system skin effect via frequency-dependent parameters line modelling and the integration of lithium-ion batteries.


\end{document}